# A Galactic Cosmic Ray Electron Spectrum at Energies from 2 MeV to 2 TeV That Fits Voyager 5-60 MeV Data at Low Energies and PAMELA and AMS-2 Data at 10 GeV Using an Electron Source Spectrum ~E$^{-2.25}$ – A Calculation Using a Monte Carlo Diffusion Model

W.R. Webber

New Mexico State University, Astronomy Department, Las Cruces, NM  88003, USA

## ABSTRACT

In this paper we fit the observed galactic cosmic ray electron spectrum from a few MeV to ~1 TeV. New data from Voyager from 5-60 MeV beyond the heliopause is used along with high energy data from the PAMELA, FERMI and AMS-2 instruments in Earth orbit. Using a Monte Carlo diffusion model for galactic propagation we obtain a source rigidity spectrum with a spectral index =-2.25 independent of energy below 10 GeV, possibly steepening above 10 GeV to ~2.40 at the highest energies. This spectrum will fit the electron data over 5 orders of magnitude to within $\pm$ 10% at both low and high energies. This steepening of the electron source spectrum could be an important feature of the acceleration process, e.g., synchrotron loss during acceleration could steepen the source spectrum. This fit requires only a single break in the rigidity dependence of the diffusion coefficient by ~1.0 power in the exponent at about 1.0 GV. The calculations also predict the distribution of electrons perpendicular to the galactic disk. The galaxy does act as a calorimeter for electrons in the energy range 0.1-10 GeV where ~80% of the electrons are trapped. At higher energies the electrons escape the galactic disk rapidly. At energies ~1 TEV the electrons loose most of their energy by synchrotron and inverse Compton emission within $\leq$ 0.5 Kpc of their origin near the galactic equator. At energies $\leq$ 0.1 GeV, because of the increase in the diffusion coefficient below ~1 GV, and the absence of other significant loss processes, the electrons escape and form a difficult to detect, but important galactic halo population.

## Introduction

The recent measurements of the cosmic ray electron spectrum at energies between 5-60 MeV beyond the heliopause by Voyager (Stone, et al., 2013) have provided a new understanding of the propagation of these lower energy electrons in the galaxy (Webber and Higbie, 2013). In this paper we extend the lower energy diffusive paradigm for electron propagation described in the above paper to higher energies, up to 1 TeV and above where synchrotron and inverse Compton losses become increasingly important. The propagation model is a Monte Carlo Diffusion Model described in an earlier paper (Webber and Higbie, 2008). The electron spectrum derived from this model is normalized to the PAMELA intensity at 10 GeV (Adriani, et al, 2011) +10% to allow for solar modulation. The important parameters in the model are the source spectra, $j(P)$, which is determined to be $P^{-2.25}$ and a diffusion coefficient $K(P) \sim P^{0.5}$ above a lower rigidity which is $\sim 1$ GV and equal to $2.5 \times 10^{28}$ $cm^2 \cdot s$ at 1 GV. These values and a boundary at 1.5 Kpc lead to a lifetime for electrons at 1 GV = $1.2 \times 10^7$ years. This lifetime is equivalent to the CR nuclei lifetime measured from 10 Be decay which, according to Yanasuk, et al., 2001, using data from the ACE experiment, is = $1.5 \times 10^7$ years at $\sim 1.0$-1.5 GV. The electrons are injected near the galactic plane at Z=0 in keeping with a possible galactic origin in SN remnants. Below $\sim 1.0$ GV the diffusion coefficient becomes $\sim P^{-0.5}$, a change of 1.0 power in the exponent.

The dominant energy loss terms in this diffusion calculation above $\sim 10$ GeV are synchrotron and inverse Compton emission both of which have an energy dependence which is $\sim E^2$. This, along with the diffusion escape of the electrons from the galaxy, results in spectra at high energies which are $\sim 0.8$ - 1.0 in the exponent steeper than the original source spectra thus leading directly to the observed high energy spectra which have exponents $\sim P^{-3.0}$ to $P^{-3.2}$, using source spectra of $P^{-2.20}$ to $P^{-2.40}$ which are typically used in the model.

At energies less than 1 GeV where diffusion dominates, the diffusion coefficient becomes $\sim K(P) \sim P^{-0.5}$ thus leading to a predicted electron spectrum which is $\sim E^{-1.5}$, the same spectrum that is measured on Voyager (Stone, et al., 2013) and about 0.75 power flatter than an $E^{-2.25}$ source spectrum.

We note that our approach of adjusting the diffusion coefficient only to interpret the complete and complex electron spectrum from a few MeV to a few TeV is different than that of Strong, et al., 2011, for example, who considered three different injection spectra, one below ~4 GV which has a spectral index = -1.6 or -2.0, one between 4 and 50 GV with a spectral index = -2.5 and one above 50 GeV with a spectral index = -2.2.

As noted below, in order to fit all of the current data from Voyager to AMS-2 we use a single constant spectral index = -2.25 between 0.01 and ~10 GeV, possibly steepening to between -2.35-2.40 at higher energies due to synchrotron energy losses during the acceleration process.

**The Data Below a Few GeV**

At energies below a few GeV the electron spectral exponent must rapidly decrease to fit the new Voyager data (Webber and Higbie, 2013; Bisschoff and Potgieter, 2014). At these lower energies the electron spectral exponent is measured to be ~1.5 $\pm$ 0.1 (Stone, et al., 2013). Thus the spectral exponent of any source spectrum for electrons must decrease from a source spectral index of -2.25, by 0.75 power below 1-2 GeV, for example. We have earlier considered that this changing electron spectrum is caused by a change in the rigidity dependence of the diffusion coefficient, K, (Webber and Higbie, 2008), following the work of Ptuskin, et al., 2006, who described the theoretical basis for such a break. Below 1 GV this rigidity dependence of K is taken to be $\sim P^{-0.5}$. This change in rigidity dependence by 1.0 power in the exponent is also consistent with the newly measured cosmic ray nuclei spectra from H to Fe measured at Voyager (Webber, 2015).

The strongest motivation for a rapid change in the value of the rigidity dependence of the diffusion coefficient comes from the electron data itself, however. The Voyager data in 2012 (Stone, et al., 2013) and the PAMELA data in 2009 (Bozio, et al., 2014), are shown in Figure 1. In each case the intensities are multiplied by $E^2$. Also shown is the interstellar spectrum obtained starting from a source spectrum of constant spectral index. It is seen from this figure that the spectral index of the interstellar spectrum has a value of 2.0 at ~0.5 GeV where the j x $E^2$ intensities are at a maximum. To determine this spectrum we have calculated the interstellar electron spectrum that will fit the 5-60 MeV Voyager data <u>and</u> the PAMELA data at 10 GeV where the solar modulation is small starting with a given source spectrum. The total intensity of electrons changes by a factor of $1.4 \times 10^6$ between 10 MeV and 10 GeV. A simple line connecting these points with this ratio would have a slope ~-2.04. To determine the actual source spectrum that would provide this ratio and also the reference intensities at 10 MeV and 10 GeV; (1) the exponent of source spectrum is varied from ~2.20 to 2.30, and (2) the rigidity of the break in the rigidity dependence of K, $P_0$, is varied from 0.316 to 1.76 GV. An excellent fit to the Voyager intensities at low energies which has a spectrum $\sim E^{-1.50}$ and the PAMELA intensity at 10 GeV with a spectrum of 3.10 at that energy, is obtained for primary electron spectra with exponent equal to ~-2.25 $\pm$0.03 and with a break at between $P_0$ = 1.00-1.33.

The actual dependence of K(P) that is used below ~1 GV, which in this case is taken to be $\sim P^{-0.5}$, will make a large difference in the low energy electron intensities and also in the spectral shape that is calculated, since the increasing value of K with decreasing energy means that the lower energy electrons will rapidly escape from the galaxy. Such a break in the rigidity range from ~1-2 GV has been predicted earlier by Ptuskin, et al., 2006, as we have noted, so this is a confirmation of this earlier work.

This procedure to determine the LIS electron spectrum is also discussed in Webber, et al., 2015, where it is used to determine the amount of solar modulation between the Voyager/Interstellar spectrum and that measured in 2009 by PAMELA at the Earth between 80 MeV and 10 GeV. The amount of this solar modulation for electrons is very sensitive to the ratio of the estimated LIS electron spectrum to the spectrum measured by PAMELA, especially in the energy range of ~1 GeV and above. This modulation data supports the LIS electron spectrum we have derived and the distinct features of the break in the P dependence of the diffusion coefficient at about 1.0 GV. Note, as seen from Figure 1, that the exponent of the calculated interstellar electron spectrum changes continuously from ~1.5 at 0.1 GV and below to ~3.0 at 10 GV with no change in source exponent. The slope of the calculated spectrum below ~0.1 GV depends strongly on the P dependence of K at these low rigidities. A slope of -1.5 is obtained only for K $\sim P^{-0.5 \pm 0.1}$. Only propagation effects are involved in the calculation; the changing dependence of K on P at lower rigidities and synchrotron and inverse Compton losses at energies greater than a few GeV.

### The Data Above 10 GeV

Here we show in Figure 2A electron (e- + e+) data over a broader energy range between 10 MeV and 2 TeV. This includes Voyager data at low energies and data from the high resolution, high statistics measurements from PAMELA (Adriani, et al., 2011), FERMI, (Ackermann, et al., 2012) and AMS-2 (Aguilar, et al., 2014) at higher energies. These spectra are plotted in a representation where a $j \sim P^{-3.0}$ spectrum is flat. A reference LIS spectrum above 10 GeV is taken to be the AMS spectrum which has an overall accuracy within $\pm$ 3%. This spectrum is shown by the red curve in Figure 2A. Also shown are the PAMELA data (blue data points). At 10 GeV the reference intensity x $E^3$ (GeV) is taken to be 220 e/m$^2$·sr·s·GeV, $\pm$5%. This reference electron intensity, x $E^{3.0}$, decreases slowly at higher energies. Between ~30 to 600 GeV this reference spectrum has an average exponent ~3.18 as noted by Aguilar, et al., 2014.

Figure 2A also shows the propagation calculations for the overall electron spectrum from 2 MV to 2 TV using the galactic propagation MCDM used earlier by Webber and Higbie, 2008, and used herein to calculate the low energy spectra with the propagation parameters as noted below. These calculated spectra are shown in figures 2A and 2B for source spectra of -2.25 (the spectrum that fits the data below 10 GeV) and in Figure 2B for source spectra of -2.30, -2.35 and -2.40 with the source exponent constant with energy throughout the entire energy range in all cases. The calculated intensities of these spectra are normalized at 1.0 GV. A comparable spectrum from Strong, et al., 2011, for source spectral indices of -2.0/-2.5/-2.2 is shown in Figure 2A as a purple line.

### The Propagation Calculations

In the MCDM each of the $10^4$ particles that are injected at each of the 64 energies between 2 MeV and 1 TeV at Z=0 could have taken up to $10^4$ equal time steps during its lifetime. This time is greater than 2 times the average lifetime of ~1.5 x $10^{14}$ sec at 1.5 GV which is the longest average lifetime observed for electrons. So the electron distributions at each energy reach their equilibrium value during the calculations.

The diffusion lifetime, t =$L^2$/K for electrons, is fixed at 1.5 x $10^7$ years at ~1.5 GV for a boundary L=$\pm$ 1.5 Kpc for the calculations. This (along with K=2x$10^{28}$/cm$^2$·s necessary for this lifetime at 1 GV) leads to a diffusion coefficient at 1 TeV of 6 x $10^{29}$ cm$^2$·sec for a dependence of K(P) ~$P^{0.5}$ above 1 GV that we have taken. This along with a boundary at $\pm$ 1.5 Kpc means that 1 TeV electrons, for example, have taken an average ~200 "steps" to reach the boundary before leaving the galaxy.

At lower energies the diffusion coefficient will become larger as well, also becoming ~ 6x$10^{29}$ cm$^2$·s$^{-1}$ at 1 MeV.

The energy dependence of the various loss terms used in the calculation is shown in Figure 3. Above a few GeV the synchrotron and inverse Compton losses, which are ~$E^2$, become increasingly important. They dominate over diffusion loss which is the largest loss term below 5-10 GeV. At these energies then, one might expect a change of slope of the observed spectrum and at also the calculated spectrum even for a constant source spectral exponent.

Most of the loss of electrons from the higher energy part of the spectrum above 100 GeV due to these losses, occurs near the galactic plane. This is because of the exponential Z dependence of the magnetic field B ~$\exp^{Z/Z_B}$ where $Z_B$ = 1.5 Kpc and the relevant photon intensities that enter into the inverse Compton calculations.

The propagation parameters we use and the model, itself, are similar to those used by Kobayashi, et al., 2004, in their two and three dimensional models. Our propagated electron spectra are also very similar to theirs, as apart from a normalization factor ~1.2.

Between ~30 GeV and 1 TeV the calculated electron spectra exponent obtained with a source exponent of -2.35 matches the reference spectral slope of -3.18 (Aguilar, et al., 2014). Below 10 GeV the average exponent has been determined to be -2.25 so the source exponent may change by at least ~0.10 above 10 GeV. If this is a continuous change starting at ~10 GeV then it could amount to ~0.05 in the exponent per decade of energy. In other words the source spectrum that leads to the observed electron spectrum has a nearly constant exponent at energies below ~10 GeV with a possible increase in the exponent from -2.25 to -2.35 with increasing energy above 10 GeV up to ~1 TeV.

## **Electron Distributions Perpendicular to the Disk of the Galaxy and the Escape of Electrons from the Galaxy**

The MCDM also allows us to determine the Z distribution of the electrons that remain in the galaxy as well as the number escaping as a function of energy. In essence, the galaxy acts like a calorimeter for the electrons as noted by Strong, et al., 2010, but, in fact, this occurs only at certain energies. In Figure 4 we show the distribution of electrons remaining in the galaxy as a function of various energies for a diffusion lifetime, $L^2/K = 1.5x10^7$ years at 1.2 GV and for the K(P) and B(r) dependences as noted above. The values of dE/dt for electrons from synchrotron loss, the inverse Compton effects and Bremsstrahlung are those shown in Figure 3.

At 1 TeV only 100 particles, or 1%, of the $10^4$ electrons that are injected at Z=0 remain in the galaxy after $10^4$ equal time steps. At 100 GeV this number is ~1200 and at 10 GeV the number is 3800. At the highest energies the "lost" particles do not escape from the galaxy but instead energy loss depletes the spectrum at Z distances <0.1-0.2 $Z_B$.

At intermediate energies between ~0.1 and 10 GeV, the galaxy does, in fact, become a diffusive calorimeter, with less than 25% of the originally injected particles actually escaping from the disk.

At energies < 0.1 GeV, however, the electrons more rapidly escape from the disk of the galaxy by diffusion, but in this case the energy loss is small. In fact, at 30 MeV and below, over 80% of the electrons originally injected actually escape from the disk of the galaxy with only a small energy loss. So in effect, our galactic disk is a copious source of electrons injected into the galactic halo at those low energies. Presumably this is true for other similar galaxies as well. This source of energetic electrons could be much larger for energetic radio galaxies. This is a new aspect of our calculations brought about by the increasing diffusion coefficient at low energies used to fit the new Voyager data.

**Summary and Conclusions**

The Voyager 1 measurements of the LIS spectrum of electrons between 5-60 MeV along with recent measurements of this electron spectrum up to and above 1 TeV by several spacecraft experiments including PAMELA, FERMI and AMS-2 have defined the galactic electron spectrum as follows: At energies below 0.1 GeV this spectrum has an exponent ~-1.5. This exponent increases with increasing energy. This change in exponent is particularly rapid at ~1 GeV where the observed spectrum is $\sim E^{-2.25}$, the same exponent as the source spectrum. At 10 GeV the observed exponent is ~-3.0 and at energies ~30 GeV and above the measured spectral exponent approaches a value of -3.18.

We have shown in this paper that the changing interstellar spectral index that is observed below about 10 GV can be viewed as a result of rigidity dependent propagation effects in a diffusing galaxy with energy loss by synchrotron and inverse Compton radiation. The source spectral exponent itself remains essentially constant as a function of energy at a value of -2.25 below 10 GeV. These propagation effects include a rapid change in the rigidity dependence of the diffusion coefficient by a power ~1.0 in the exponent at about 1 GV. Using this variation of the diffusion coefficient and also using a single "source" spectral index of -2.25 we are able to reproduce the LIS electron intensities observed at both 10 MeV by Voyager and at 10 GeV by PAMELA to within 10%, as well as the slope of the observed spectrum changing from $E^{-1.50}$ below ~100 MeV to $E^{-3.0}$ at 10 GeV.

Above ~10 GeV it appears from a comparison of PAMELA, FERMI and AMS-2 data, which within themselves, appear to agree to within $\pm$ 10%, and the calculated spectra, that the source spectral index may increase with increasing energy, e.g., from -2.25 at 10 GeV to -2.35 at 1 TeV. This indicates a possible increase in the average spectral index by as much as ~0.050 per decade of energy at the highest energies. This could be possibly related to features of the acceleration process of these electrons at the highest energies, e.g., the rate of acceleration vs. the loss of energy due to the synchrotron effect in the accelerating region, thus defining features of the acceleration process.

Calculations of the distribution of electrons perpendicular to the disk illustrate that at energies between 0.1 and 10 GeV the galaxy acts as a calorimeter for the electrons produced within it along the equatorial plane. Less than 25% of these electrons escape into the galactic halo. At energies ≤0.1 GeV, however, a rapidly increasing fraction of lower energy electrons diffuse beyond the boundary of the galactic disk and escape. This is because of the increasing diffusion coefficient at low energies and the lack of any large competing loss process. So from a few MeV and up to about 100 MeV the disk of the galaxy actually becomes a copious source of electrons into the galactic halo. This is a new aspect of our studies.

Above 100 GeV most of the electrons never leave the disk of the galaxy, losing energy initially at a rapid rate near the galactic equatorial plane. As a result only a few percent of those injected above 100 GeV even remain beyond a Z=0.2-0.3 x $Z_B$ (~300-500 pc) from the galactic plane source.

**<u>Acknowledgments:</u>** The Author appreciates discussions with Paul Higbie. This paper would not have been possible without the assistance of Tina Villa.

# REFERENCES

Ackermann, M., et al., 2012, Phys. Rev. Lett., 108, 011103

Adriani, O., Barbarino, G. C., Bazilevskaya, G.A., et al. 2011, Phys. Rev. Lett., 106, 201101-1

Adriani, O., Barbarino, G. C., Bazilevskaya, G.A., et al. 2015, (to be published)

Aguilar, M., Alberti, G. and Alpat, B., 2013, Phys. Rev. Lett., 110, 141102

Bisschoff, D. and Potgieter, M.S., 2014, Ap.J., 794, 166

Kobahashi, T., et al., 1999, Proc. 26th ICRC, 3, 61

Kobahashi, T., Komori, Y., Yoshida, K. and Nishimura, J., 2004 Ap.J., 601, 340

Ptuskin, V.S., Moskalenko, I.V., Jones, F.C., et al., 2006, Ap.J., 642, 902-916

Stone, E. C., Cummings, A. C., McDonald F.B., et al. 2013, Sci, 341, 150-153

Strong, A.W., Porter, T.A., Digel, S.W., et al., 2010, Ap.J., 722, L58

Strong, A.W., Orlando, E. and Jaffe, T.R., 2011, A&A, 534, A54

Webber, W.R. and Rockstroh, J.M., 1997, Adv. Space Res., 19, 817

Webber, W. R. and Higbie, P. R. 2008, JGR, 113, A11106

Webber, W.R. and Higbie, P.R., 2013, http://arXiv.org/abs/1308.6598

Webber, W.R., 2015, http://arXiv.org/abs/1508.01542

Webber, W.R., Potgieter, M.S. and Bozio, M., 2015 (to be published)

Yanasak, N.E. Wiedenbeck, M.E. and Mewaldt, R.A., 2001, Ap.J., 563, 768

# FIGURE CAPTIONS

**Figure 1:** This figure is in a j x$E^2$ vs. E format. Measurements (in red) of the GCR electron spectrum between 5-60 MeV made by V1 beyond the heliopause at ~122 AU (Stone, et al., 2013), and those made by PAMELA in 2009 up to 10 GeV and above (Bozio, et al., 2014), at a time of minimum solar modulation at the Earth (in blue) are shown. The black vertical line at 60 MeV would represent a solar modulation factor = 500. The black line is the calculated interstellar spectrum using a Monte Carlo diffusion model as described in this paper.

**Figure 2A:** This figure is in a j x $E^3$ vs. E format. It includes the electron data from Voyager experiment (Stone, et al., 2013), the PAMELA experiment (Adriani, et al., 2011), the FERMI experiment (Ackermann, et al., 2012) and also from AMS-2 (Aguilar, et al., 2014). The black lines are the spectra obtained using source spectra of $E^{-2.25}$ and $E^{-2.35}$ normalized at 1 GV.

**Figure 2B:** A figure in a j x $E^3$ vs. E format showing the galactic electron spectrum above 10 GeV measured by AMS-2, FERMI and PAMELA. Calculated electron spectra for source spectra ~ $P^{-2.25}$, $P^{-2.30}$, $P^{-2.35}$ and $P^{-2.40}$ independent of energy and with a diffusion coefficient K ~ $P^{0.5}$ above 1 GeV are shown. The $P^{-2.25}$ spectrum is normalized to 2.20 electrons/m$^2$·sr·s·GeV at 10 GeV. All spectra are normalized at 1 GeV. The high energy points from Kobayshi, et al., 1994 are shown as x's with error bars.

**Figure 3:** The various energy loss or escape processes affecting the spectrum of GCR electrons during their propagation in the galaxy. The line labelled sync + IC represents the sum of the synchrotron (B=5μG) and inverse Compton loss, both of which are ~$E^2$. The line labelled Br is the Bremsstrahlung loss which is ~E and ion is the ionization loss which is almost independent of energy. The red lines represents characteristic times for diffusion escape of electrons out of the galaxy. This lifetime is "normalized" to the E/(dE/dt) lifetime of the energy loss processes by using the measured lifetime of cosmic ray nuclei of $1.5 \times 10^7$ yr. at 1.2 GV, as described by Yanasuk, et al., 2001.

**Figure 4:** Distribution of events perpendicular to the galactic disk, Z=0. The plot shows the number of electrons remaining per 0.1 $Z_B$ interval for each $10^4$ injected, with $Z_B$=1.5 Kpc.

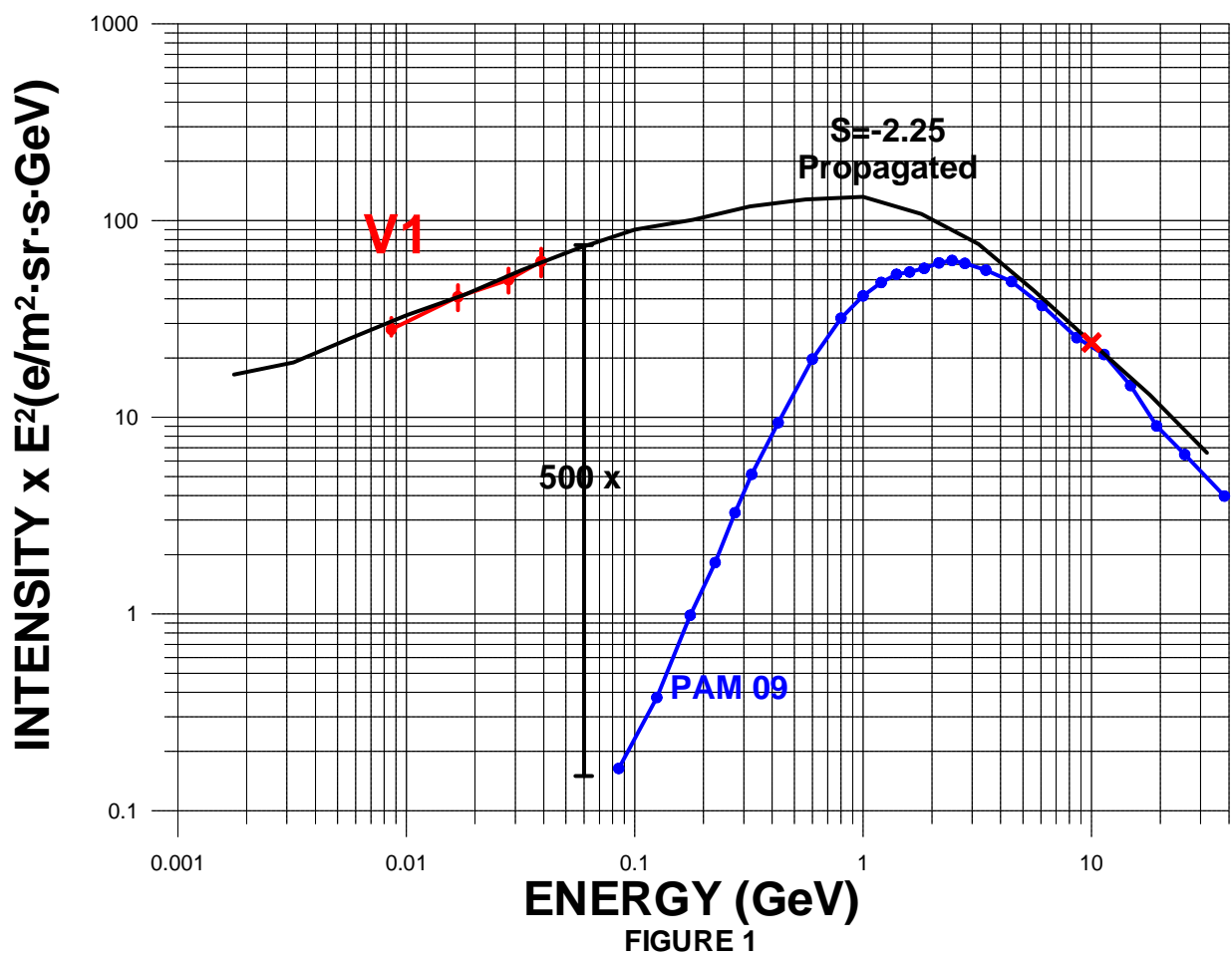

FIGURE 1

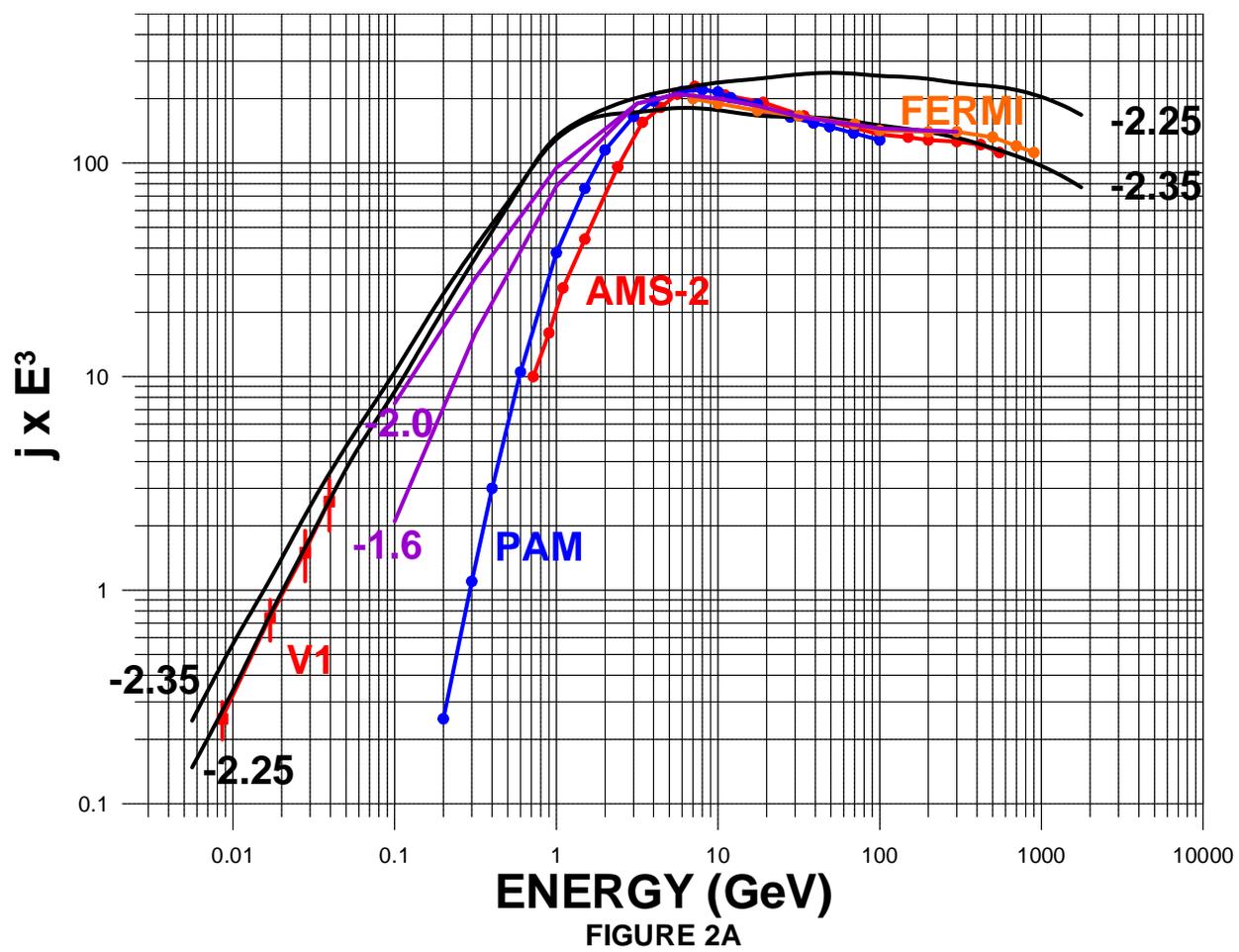

**FIGURE 2A**

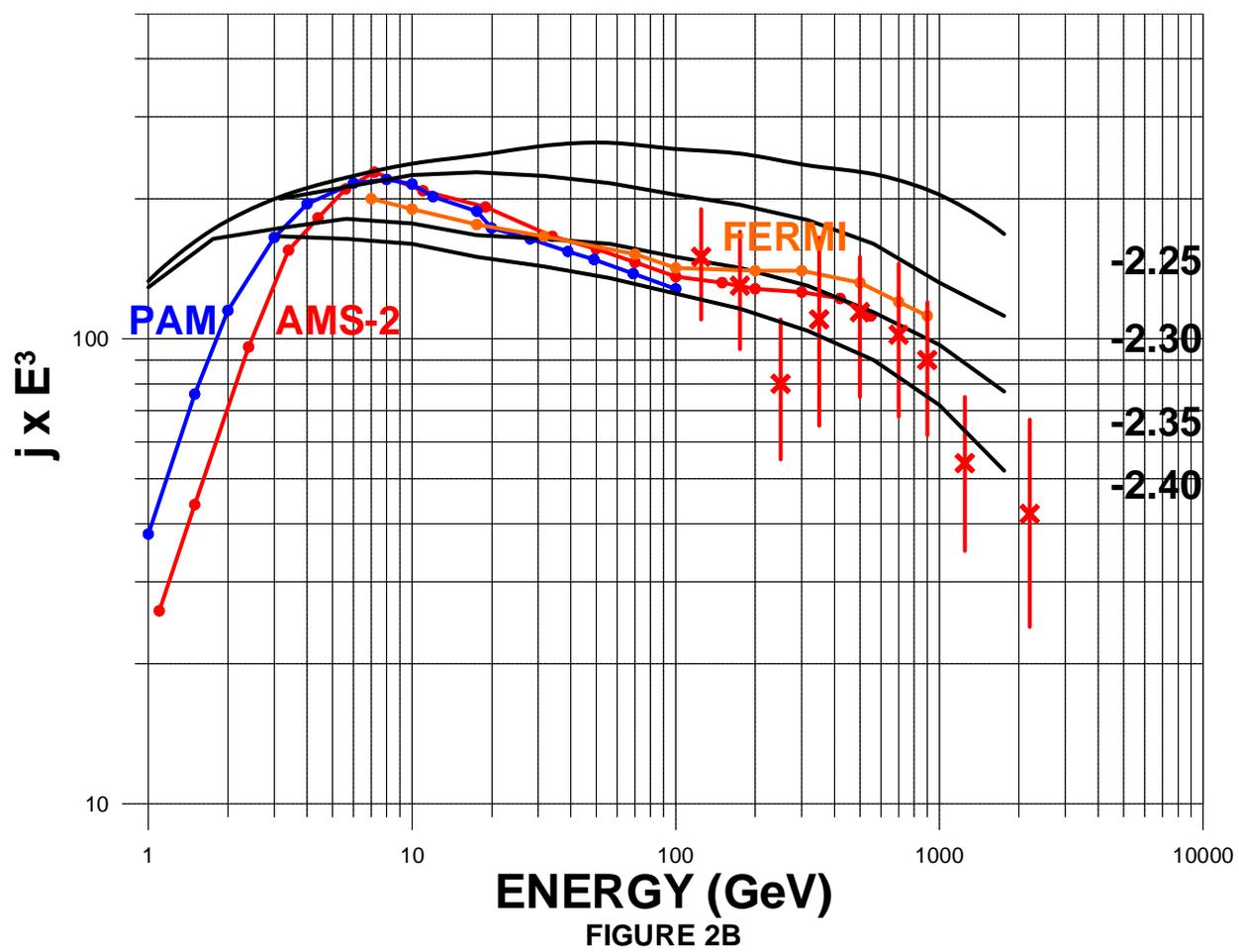

**FIGURE 2B**

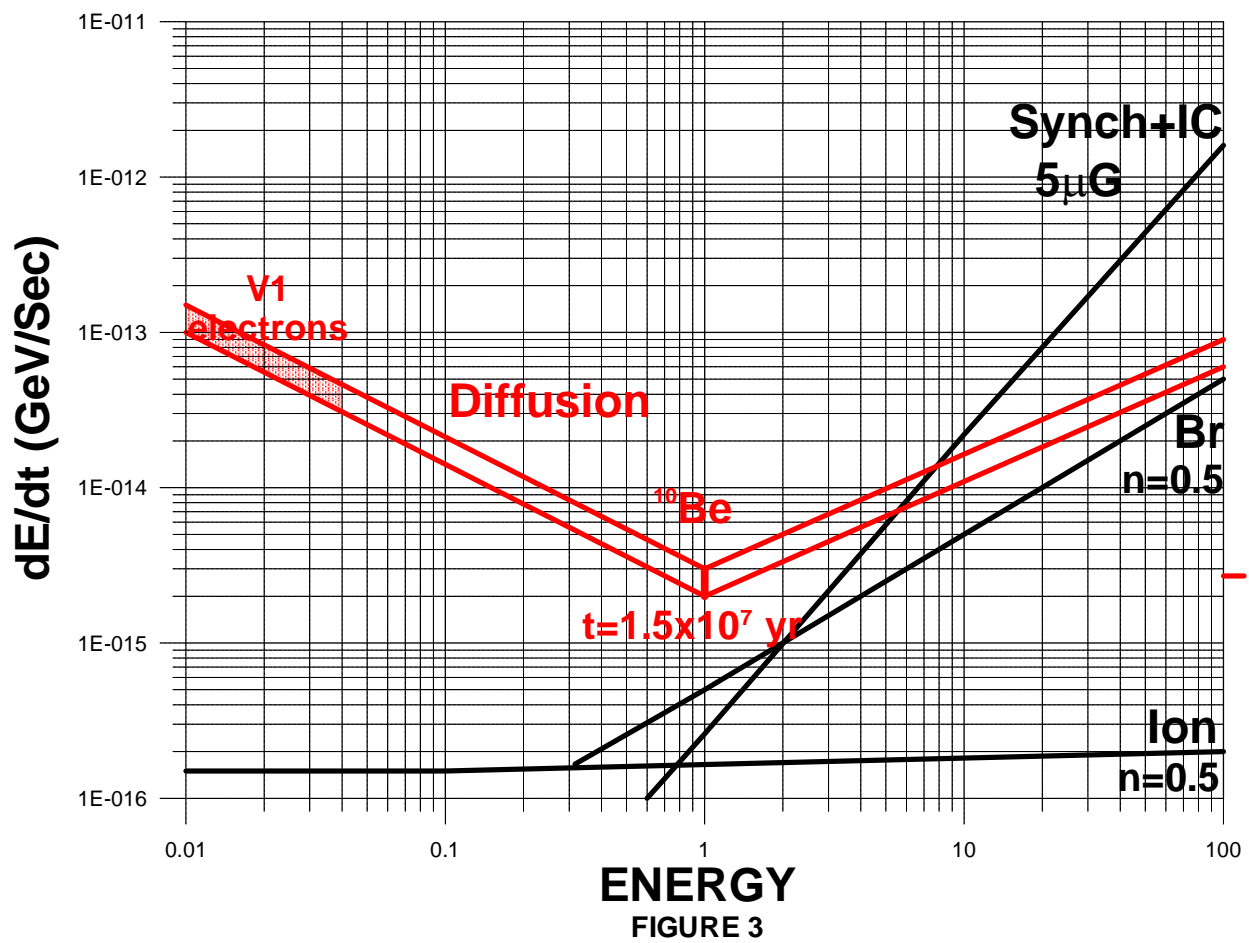

**FIGURE 3**

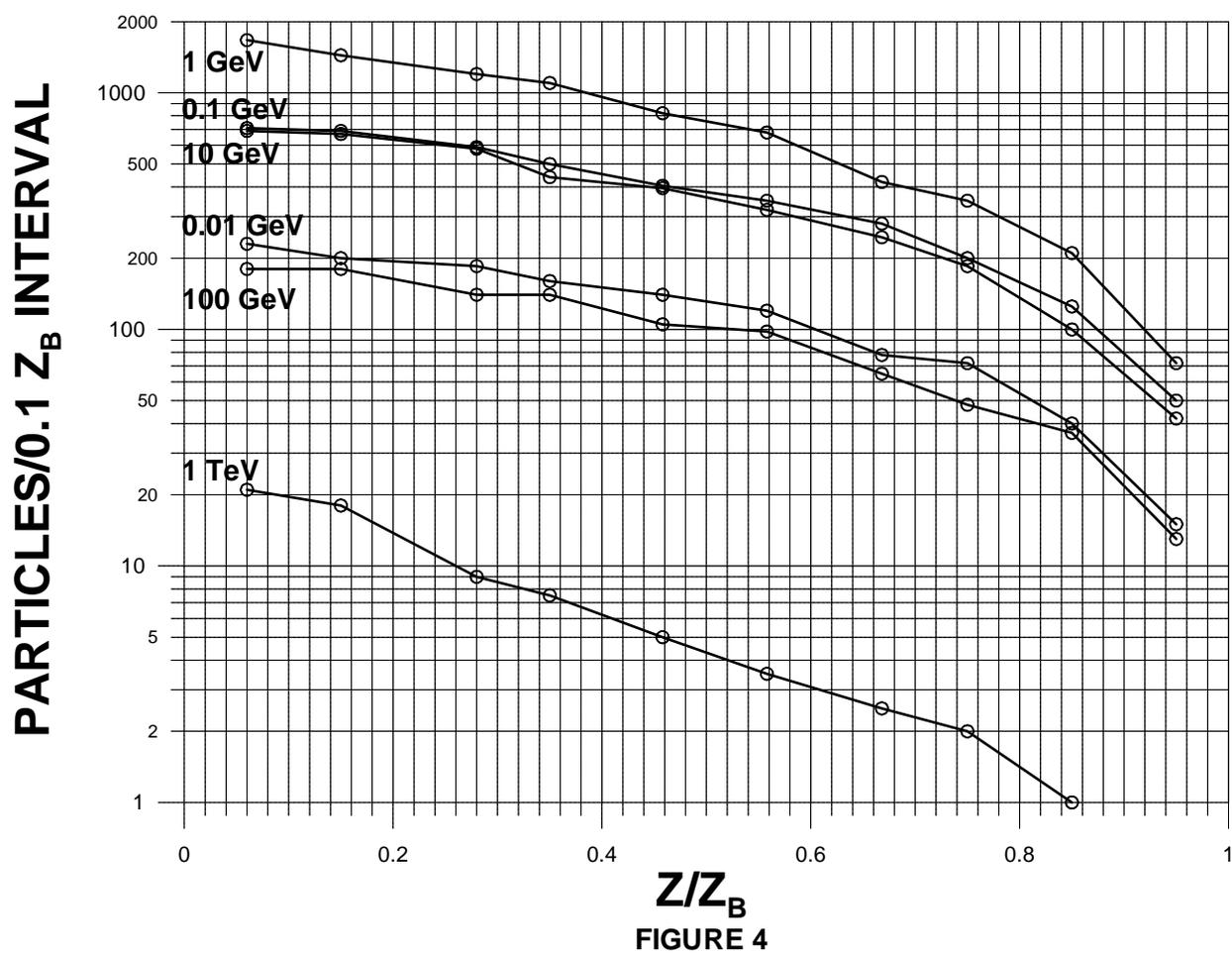

**FIGURE 4**